\documentclass[12pt,epsf,amssymb]{article}

\usepackage{graphicx}
\usepackage{axodraw}
\usepackage{amsfonts}

\begin{document}
\baselineskip 0.6cm
\newcommand{\gsim}{ \mathop{}_{\textstyle \sim}^{\textstyle >} }
\newcommand{\lsim}{ \mathop{}_{\textstyle \sim}^{\textstyle <} }
\newcommand{\vev}[1]{ \left\langle {#1} \right\rangle }
\newcommand{\bra}[1]{ \langle {#1} | }
\newcommand{\ket}[1]{ | {#1} \rangle }
\newcommand{\Dsl}{\mbox{\ooalign{\hfil/\hfil\crcr$D$}}}
\newcommand{\nequiv}{\mbox{\ooalign{\hfil/\hfil\crcr$\equiv$}}}
\newcommand{\nsupset}{\mbox{\ooalign{\hfil/\hfil\crcr$\supset$}}}
\newcommand{\nni}{\mbox{\ooalign{\hfil/\hfil\crcr$\ni$}}}
\newcommand{\EV}{ {\rm eV} }
\newcommand{\KEV}{ {\rm keV} }
\newcommand{\MEV}{ {\rm MeV} }
\newcommand{\GEV}{ {\rm GeV} }
\newcommand{\TEV}{ {\rm TeV} }

\def\diag{\mathop{\rm diag}\nolimits}
\def\tr{\mathop{\rm tr}}

\def\Z{\mathbb{Z}}
\def\R{\mathbb{R}}
\def\C{\mathbb{C}}
\def\Im{\mathrm{Im}}
\def\Re{\mathrm{Re}}
\def\vol{\mathrm{vol}}

\def\Spin{\mathop{\rm Spin}}
\def\SO{\mathop{\rm SO}}
\def\O{\mathop{\rm O}}
\def\SU{\mathop{\rm SU}}
\def\U{\mathop{\rm U}}
\def\Sp{\mathop{\rm Sp}}
\def\SL{\mathop{\rm SL}}


\begin{titlepage}

\begin{flushright}
LBNL-54530 \\
UT04-05 \\
\end{flushright}

\vskip 2cm
\begin{center}
{\large \bf GUT Phase Transition and Hybrid Inflation}

\vskip 1.2cm
Taizan Watari${}^a$ and T.~Yanagida${}^b$

\vskip 0.4cm
${}^a$ 
{\it Department of Physics, University of California at Berkeley,\\ 
          Berkeley, CA 94720, USA} \\
${}^b$
{\it Department of Physics, University of Tokyo, \\
          Tokyo 113-0033, Japan}\\
\vskip 1.5cm
\abstract{The supersymmetric model of hybrid inflation is 
interesting not only because of its naturalness 
but also because of another reason. Its energy scale 
determined by the COBE normalization is $10^{15}\mbox{--}10^{16}$ 
GeV. It happens to be the energy scale of interest 
in particle physics, namely, the mass scale of right-handed 
neutrinos or the energy scale of the gauge-coupling unification. 
It is true that topological defects are produced 
after the hybrid inflation if it is related to a U(1)$_{B-L}$ or 
a GUT-symmetry breaking, and hence one cannot naively identify 
models of particle physics with that of inflation.
But those defects are not necessarily found in modified models. 
We show in this article that quite a simple extension 
of the minimal supersymmetric hybrid inflation model is free from 
monopoles or cosmic strings. 
Moreover, it happens to be exactly the same 
as a well-motivated extension of SU(5)-unified theories.  
The vacuum energy is dominated by F-term. 
The $\eta$-problem is not necessarily serious when 
the model is realized by D-branes. 
Although it has been considered that a coupling constant has to be 
very small when the vacuum energy is dominated by F-term, 
this constraint is not applied either to the D-brane model. 
They are due to a particular form of the K\"{a}hler potential and
 interaction of the model.
Reheating process is also discussed.
}

\end{center}
\end{titlepage}


Inflation explains naturally the initial conditions 
of the standard big-bang cosmology \cite{Guth}. 
The universe had experienced superluminal expansion due to 
vacuum energy, before it underwent phase transition and 
entered into the radiation dominated 
Friedmann--Robertson--Walker universe.
The original model \cite{Guth} was associated with 
the first order phase transition of grand unified models 
of elementary particles.
The idea was extremely attractive because the well-motivated model  
of particle physics was considered to provide a solution 
to serious problems in cosmology.

However, the inflation associated with the first order 
phase transition (referred to as the old inflation model) 
does not work \cite{Hawking,Guth2} because of its unsuccessful 
reheating process\footnote{
A possible way out of this problem is proposed recently in 
\cite{Dvali-Kachru}.}.
An alternative to the old inflation is the slow-roll inflation 
\cite{new}. 
The positive vacuum energy lasts for sufficient amount of time 
not because the inflaton field is trapped at a local minimum 
of its potential, but because it rolls down the potential very slowly.
The potential has to be extremely flat for this to happen.
One of the important virtues of the slow-roll inflation is that 
it produces the scale-invariant density perturbation through 
the quantum fluctuation of the inflaton field.

It is far from natural, however, that there exists such a flat 
potential. Theorists have been working for decades to 
search for a natural model of inflation.
The hybrid inflation model \cite{LindeH} is most promising 
among many trials. 
It has a natural and simple extension with supersymmetry (SUSY),  
and hence the flatness of the potential is preserved (to some extent) 
by radiative corrections.
The minimal SUSY extension consists of three ${\cal N}$ = 1 SUSY chiral
multiplets $\Phi$, $\Psi$ and $\bar{\Psi}$, and one U(1) vector  
multiplet $V$. $\Psi$ and $\bar{\Psi}$ are charged under the U(1) 
by $\pm 1$, while $\Phi$ is neutral. The superpotential is  
\cite{Copeland,DSS,Dterm}
\begin{equation}
 W = \sqrt{2} \lambda \Phi (\Psi \bar{\Psi} - \zeta), 
\label{eq:hybrid}
\end{equation}
and there may be a Fayet--Iliopoulos D-term ${\cal L} = - \xi D_V$.
The inflaton is $\Phi$. While its value 
is large, $\Psi$ and $\bar{\Psi}$ are at the origin, and the
vacuum energy is $\rho_{cos} = v_0^4 \equiv 
(1/2) (|\lambda (2 \zeta)|^2 + g^2 \xi^2 )$, where $g$ is the gauge 
coupling constant of the U(1) vector field.
When the value of $\Phi$ becomes sufficiently small, $\Psi$ and 
$\bar{\Psi}$ begins to be non-zero, and the slow-roll 
inflation comes to an end.
The U(1) symmetry is broken down spontaneously, 
and the decay process of $V,\Phi,\Psi$ and $\bar{\Psi}$ reheats 
the universe.
The potential of the inflaton $\Phi$ is flat \cite{Copeland} 
even after supergravity corrections are taken into account 
(under certain assumptions, as we review later). 
Thus, this is a remarkably successful model.

Now, how is this model related to models of particle physics?
The estimate of the energy scale $\sqrt{\zeta}$ or $\sqrt{\xi}$ 
is obtained by 
imposing the COBE normalization on the density perturbation, 
\begin{equation}
 \frac{\delta \rho}{\rho} = 
  \frac{1}{\sqrt{75} \pi}\frac{V^{\frac{3}{2}}}{M_{\rm pl}^3 V'} 
 \simeq 1.9 \times 10^{-5},
\end{equation}
and assuming certain amount of e-fold number. 
It is roughly $10^{15}\mbox{--}10^{16}$ GeV \cite{LR,AHKY}. 
It happens to be of theoretical interest from the 
point of view of particle physics. Namely, it is near 
the mass scale of right-handed neutrinos or 
the energy scale of unified theories. 
This coincidence has led many people to consider that 
the phase transition may be related to the symmetry breaking 
of U(1)$_{B-L}$ or unified theories, where $\Psi$ and $\bar{\Psi}$ 
breaks those symmetries \cite{DSS}. 
Difficulties in this direction are topological defects 
produced after the phase transition.
If the phase transition is associated with 
the SU(5)$_{\rm GUT} \rightarrow$ 
SU(3)$_C \times$SU(2)$_L \times$U(1)$_Y$, 
the GUT monopoles are created too much, in general.
The above phase transition 
cannot be preceded by the inflation.
This means that the simple SU(5) unification has to be abandoned, 
if the hybrid inflation were to be identified with the latest stage 
of the cascade break down of a large unified symmetry.  
If the minimal SUSY model is identified with the model 
of U(1)$_{B-L}$-symmetry breaking, the cosmic strings 
are created too much; 
$\lambda \lsim 10^{-4}$ is required \cite{cos-strA,cos-strB} 
to render the CMB distortion sufficiently small\footnote{ 
The constraint in the text is obtained in the case 
where the vacuum energy comes only from the D-term.
It is loosened when the vacuum energy 
is dominated by F-term; this is because 
the tension of the cosmic string scales as $(\lambda/g) \zeta$, 
rather than $\zeta$ or $\xi$.
The constraint will be $\lambda^2/g \lsim 10^{-4}$.
Thus, the constraint is still stringent. {\it c.f.} \cite{cos-strC}, 
where the constraint in the F-term dominated case is discussed with 
an assumption $\lambda = g$.}.

The topological defects, however, are not inevitable predictions 
of unified theories or hybrid inflation.  
The topological defects associated with a symmetry breaking 
$G \longrightarrow H$ are characterized by homotopy groups; 
cosmic strings by $\pi_1(G/H)$ and monopoles by $\pi_2(G/H)$.
They are determined by examining the homotopy exact sequence,  
\begin{equation}
 0 \rightarrow \pi_2(G/H) \rightarrow \pi_1(H) \rightarrow 
               \pi_1(G) \rightarrow \pi_1(G/H) \rightarrow 0.
\end{equation}
When $G$ contains U(1), as in the minimal SUSY model 
of hybrid inflation, $\pi_1(G/H)$ tends to be non-trivial 
and the cosmic strings to be stable. 
When $H$ contains U(1), as in the standard model, $\pi_2(G/H)$ 
tends to be non-trivial, and the monopoles tend to be 
in the spectrum. 
However, if both $G$ and $H$ contains U(1) and 
\begin{equation}
 \left(\pi_1(H) \simeq {\bf Z} \right) \rightarrow 
 \left(\pi_1(G) \simeq {\bf Z} \right) 
\end{equation}
is an isomorphism, just as in the electroweak phase transition 
$(G \simeq \SU(2)_L \times \U(1)_{Y}) \rightarrow 
(H \simeq U(1)_{QED})$, neither cosmic strings nor monopoles 
are left behind the phase transition.

The simplest extension along this line is to 
to introduce an SU(2) vector multiplet and promote 
the chiral multiplets $\Psi$ and $\bar{\Psi}$ into 
``two Higgs doublets'';
$\Psi_i$ is ({\bf 2},$+1$) of SU(2)$\times$ U(1), and 
$\bar{\Psi}^i$ ({\bf 2},$-1$).
The superpotential is 
\begin{equation}
 W = \sqrt{2} \lambda \Phi (\Psi_i \bar{\Psi}^i - \zeta).
\end{equation}
The parameter $\sqrt{\zeta}$ is of the order of 
$10^{15}\mbox{--}10^{16}$ GeV.
Since the values of $\Psi_i$ and $\bar{\Psi}^i$ are zero during 
the slow-roll stage, the extension does not affect the dynamics of 
inflation essentially\footnote{The coefficient of the 1-loop 
correction to the inflaton potential is changed.}.
When the slow-roll inflation comes to an end, the system falls down 
to the vacuum, and the symmetry is broken down to U(1).
It is remarkable that there is no vacuum moduli in this extension. 
All the particles in the spectrum have masses of the order of 
$g \sqrt{\zeta}$ or $\lambda \sqrt{\zeta}$ at the vacuum.  
Thus, this extension is free from moduli problem, and moreover, 
free from harmful relics of topological defects\footnote{A model 
based on SU(2)$\times$U(1) gauge group is found 
in \cite{DSS}. See also \cite{LP}. 
The relation between the defects and inflation is studied extensively 
in the context of grand unified theories in \cite{JRS}.
}.
One can see that the same thing happens when the model is extended 
to SU($M$)$\times$U($N$) ($M > N$) system with the superpotential 
\begin{equation}
 W = \sqrt{2} \lambda  
   ( \bar{\Psi}^i_{\;\; \alpha} \Phi \Psi^\alpha_{\;\; i}
    - \zeta \Phi ) 
   + \sqrt{2} \lambda' 
     \bar{\Psi}^i_{\;\; \alpha} (t^a)^\alpha_{\;\; \beta} \Phi^a 
     \Psi^\beta_{\;\; i}, 
\label{eq:ext2}
\end{equation}
where $\alpha, \beta = 1,\ldots,N$ and $i=1,\ldots,M$ denote SU($N$) 
and SU($M$) indices, $t^a$ ($a=1,\ldots, N^2-1$) are 
Hermitian matrices of the generators of the SU($N$) Lie algebra, 
and $\Phi^a$ are chiral multiplets. 
The gauge group is broken down 
to SU($N$)$\times$SU($M-N$)$\times$U(1) after phase transition.
It is important to notice that the Fayet--Iliopoulos D-term 
parameter $\xi$ has to vanish in this class of extensions, because 
there would be no supersymmetric vacuum otherwise.
Thus, the vacuum energy is dominated by F-term.

Incidentally, there is a model of SU(5)-unified theories\footnote{
We mean, by SU(5)-unified theories, those satisfying 
the following properties: 
a) there is a good explanation for the SU(5)-symmetric gauge-coupling 
unification, and 
b) the U(1)$_Y$ charge assignment on quarks and leptons are 
explained in an elegant way in unified multiplets. 
The grand unification (combining all the forces into a single force) 
is not required; it is sufficiet for both properties, 
but not necessary. 
Product groups are quite natural if models are embeded 
in brane world. Quantization of U(1) charges also follows naturally. }
 that has exactly the same matter contents and superpotential as (\ref{eq:ext2}) 
\cite{Izawa-Yanagida}. 
The gauge group is\footnote{There is a cousin model, whose gauge
group is SU(5)$_{\rm GUT} \times$U(2)$_{\rm H}$. Review of this model 
can be found, e.g., in \cite{Ibe-Watari}.} 
SU(5)$_{\rm GUT} \times$U(3)$_{\rm H}$, i.e., $M=5$ and $N=3$, 
at high energy, and is broken down to SU(3)$\times$SU(2)$\times$U(1), 
the standard-model gauge group.
The parameter $\sqrt{\zeta}$ provides the GUT scale, 
$\sim 10^{16}$ GeV. 
This model was obtained through discussion independent of 
cosmology and purely in the context of unified theories.
It is one of a few classes of models that address  
the two major problems of unified theories in terms of symmetry;  
the problems are the doublet--triplet splitting and 
the suppressed dimension-five proton decay. 
For more details of this model, see \cite{Izawa-Yanagida} or 
review sections in \cite{Ibe-Watari,WY04}.
Thus, the phenomenologically well-motivated model of 
SU(5)-unified theories happens to have the same superpotential 
as the hybrid inflation model,
and is free from the cosmic-string and monopole problems.
The question is whether the model is able to realize 
the {\it slow-roll} inflation or not\footnote{``Phase transition'' 
in the title of this article may be a little misleading; 
strictly speaking, the phase transition is not the slow-roll stage, 
but a process of the graceful exit from inflation.}.

Notice that the Fayet--Iliopoulos D-term parameter $\xi$ 
has to be zero in the extended model, and hence 
the vacuum energy is dominated by F-term. 
Thus, there is so called $\eta$-problem \cite{Copeland,LR}, as 
briefly explained below.
If the vacuum energy were dominated by D-term, i.e., 
$\xi \neq 0, \zeta = 0$, the inflaton potential would be \cite{Dterm}
\begin{equation}
 V(\sigma) = \frac{1}{2}g^2 \xi^2 
   \left( 1 + \frac{g^2}{4\pi^2} \ln \left( \frac{\sigma}{\sqrt{\xi}}
                                     \right)
   \right)
\label{eq:Dterm-pot}, 
\end{equation}
and is independent\footnote{It depends on the gauge kinetic term 
of $V$, instead. However, $(\Phi/M_{\rm pl})$-corrections 
are absent there, if an $R$ symmetry is imposed.} 
of the K\"{a}hler potential (kinetic term) of $\Phi$.
However, the potential becomes \cite{Copeland,LR} 
\begin{equation}
 V(\sigma) = \frac{1}{2} | \lambda 2 \zeta|^2 
    \left(1 + 
          \frac{\lambda^2}{4\pi^2} 
           \ln \left(\frac{\sigma}{\sqrt{\zeta}}\right)
         + \frac{k}{2} \left(\frac{\sigma}{M_{\rm pl}}\right)^2 
         + \cdots
    \right)
\label{eq:Fterm-pot}
\end{equation}
when $\xi =0, \zeta \neq 0$, where
\begin{equation}
 K = \Phi^\dagger \Phi 
   - \frac{k}{4}\frac{\left(\Phi^\dagger \Phi \right)^2}{M_{\rm pl}^2}
   + \cdots.
\label{eq:kahler}
\end{equation}
The slow-roll conditions are 
\begin{equation}
 \eta \equiv \frac{M_{\rm pl}^2 V''}{V} \ll 1, \qquad 
 \epsilon \equiv \frac{1}{2}
         \left(\frac{M_{\rm pl}V'}{V}\right)^2 \ll 1,
\label{eq:sr}
\end{equation}
and they are sensitive to $M_{\rm pl}$-suppressed corrections, 
such as the third term in (\ref{eq:Fterm-pot}). 
Thus, even the $M_{\rm pl}$-suppressed corrections in the K\"{a}hler 
potential has to be under control ($k \ll 1$) 
for the sufficiently flat potential.
However, the second term in the K\"{a}hler potential 
seems to have the same symmetry property as the first term, 
and it is hard to understand why the second term in (\ref{eq:kahler}) 
is absent.

References \cite{WY-hybrid} pointed out that 
the minimal SUSY hybrid inflation model with superpotential 
(\ref{eq:hybrid}) can be compatible 
with an ${\cal N}$ = 2 SUSY\footnote{The ${\cal N}$ = 2 SUSY was also 
discovered in \cite{cos-strA}.}. 
An ${\cal N}$ = 2 rigid SUSY is enhanced in the U(1) gauge theory,  
i.e., in the inflation sector, when $\lambda = g$. 
It is considered in \cite{WY-hybrid} that the enhanced SUSY
may be responsible for solving the $\eta$-problem, since 
even the K\"{a}hler potential is under strict control of the SUSY.

If the hybrid inflation model with the superpotential 
(\ref{eq:ext2}) is identified with the model of unified theories, 
the SU($M=5$)$_{\rm GUT}$ gauge interaction connects the 
particles responsible for inflation/GUT-breaking  
with quarks and leptons. 
Thus, ``the inflation/GUT-breaking sector'' is not well-defined.
Since the SU(5)$_{\rm GUT}$-{\bf adj.} chiral multiplet is not
introduced in the model, and since quarks and leptons are in the 
chiral representations, 
there is no ${\cal N}$ = 2 SUSY in the total model. 
However, ``the GUT-breaking sector'' is decoupled 
(except through gravity) from other particles, 
when SU(5)$_{\rm GUT}$ interaction and 
a couple of others are turned off \cite{IWY}. 
Thus, in this case, the corresponding statement is that 
the inflation/GUT-breaking sector is decoupled 
when some couplings are turned off, and the ${\cal N}$ = 2 rigid 
SUSY is enhanced in the decoupled sector 
when $\lambda = g$ is satisfied \cite{IWY}.
Analyses of the model in \cite{Fujii-Watari,Ibe-Watari} suggests 
that the relation $\lambda=g$ is preferred phenomenologically.

Although the inflation sector has the ${\cal N}$ = 2 rigid SUSY 
(if $\lambda = g$), quarks and leptons have only an ${\cal N}$ = 1 
SUSY.
Thus, the gravity, which combines both of them, has 
only the ${\cal N}$ = 1 SUSY. 
In particular, the model of inflation has to be, in the end, 
embedded in an effective theory in D = 4 ${\cal N}$ = 1 supergravity.
Thus, it is not easy to expect that the rigid ${\cal N}$ = 2 SUSY
plays some roles in the flatness of the potential, 
since the slow-roll conditions (\ref{eq:sr}) require 
$M_{\rm pl}$-suppressed corrections be under control. 
Reference \cite{WY-hybrid} considered that it may be 
in the following way that the $\eta$-problem is solved.
\begin{enumerate}
 \item There should be a cut-off scale $M_*$ other than $M_{\rm pl}$, 
       and $M_* < M_{\rm pl}$. 
       They are independent parameters, and 
       there is a well-defined limit which takes $M_{\rm pl}$ 
       to infinity, keeping $M_*$ finite.
 \item The inflation sector decouples completely from other sectors 
in that limit.
 \item The ${\cal N}$ = 2 rigid SUSY is enhanced in that limit in 
the inflation sector.
 \item  Flat inflaton potential is obtained because 
    the rigid ${\cal N}$ = 2 SUSY controls 
    $(\Phi/M_*)$-corrections in the K\"{a}hler potential.
 \item Finite $1/M_{\rm pl}$ effects provides the inflation sector 
only with the coupling of the particles in the sector with gravity.
In other words, the effects do not alter the inflaton potential 
directly.
\end{enumerate}
Suppose that all these assumptions are correct. 
When the finite $1/M_{\rm pl}$ is restored from the limit, 
the only effect to the inflaton potential is through loop amplitudes.
Loop factors and positive powers of $(M_*/M_{\rm pl})$ 
render the corrections small, and the potential flat.

If the inflation sector is identified with the GUT-breaking sector, 
some other interactions also have to be turned off 
in the limit in the assumption 1, and the assumption 5 
has to hold true also for the finite effects of those couplings. 

It was not clear in \cite{WY-hybrid} 
what these assumptions imply, or whether they are justified or not.
Moreover, there is an additional difficulty in cosmology 
in the presence of the ${\cal N}$ = 2 rigid SUSY; 
the spectrum in the inflation sector becomes degenerate 
if $\lambda = g$, and then the reheating does not proceed 
smoothly because of the degeneracy \cite{WY-hybrid}. 
It was also unclear how this difficulty is solved.
Embedding the model into Type IIB brane world clarifies 
the meaning of the assumptions, justifies some them, 
and solves the difficulty, as seen below.

The first progress in this direction was made in the context of 
unified theories \cite{IWY}. 
The GUT-breaking sector and its interaction fit very well to 
the D3--D7 system of the Type IIB string theory. 
The D = 4 ${\cal N}$ = 2 SUSY SU($N$)$\times$U(1) 
vector multiplets $(V^a,\Phi^a)$ and $(V,\Phi)$ 
come from D3--D3 open strings, and the hypermultiplet 
$(\Psi^\alpha_{\;\; i},\bar{\Psi}^i_{\; \; \alpha})$ from 
the D3--D7 open strings.
When the D3--D7 system is put into a Calabi--Yau 3-fold, 
the total theory has only D = 4 ${\cal N}$ = 1 SUSY. 
The low-energy effective theory should be given by D = 4 
${\cal N}$ = 1 supergravity.  
However, as long as the local geometry of a Calabi--Yau 
3-fold around the D3-brane preserves the D = 4 ${\cal N}$ = 2 SUSY 
of the D3--D7 system, 
the matter contents and their interactions are rigid ${\cal N}$ = 2 
supersymmetric (when quantum corrections are neglected) 
\cite{IWY,WY01,WY04}.
The D = 4 ${\cal N}$ = 1 SU($M=5$)$_{\rm GUT}$ vector multiplet 
comes from D7--D7 open strings.
The GUT phase transition is the process of D3--D7 bound 
state formation \cite{WY01}.

$N$ ordinary D3-branes were used for SU($N$)$\times$U(1),
and the origin of the Fayet--Iliopoulos F-term parameter $\zeta$ 
in (\ref{eq:ext2}) was identified with B-field background 
(although the zero mode is absent because of 
the orientifold projection) in \cite{IWY}. 
$N$ fractional D3-branes on ${\bf C}^2/{\bf Z}_M \times {\bf C}$ 
were used instead in \cite{WY01}, so that 
unwanted moduli (in Higgs branch) are removed. 
We thought that the vacuum expectation value (VEV) of 
twisted NS--NS sector fields may provide the 
non-vanishing Fayet--Iliopoulos parameter \cite{WY01}.
However, the U(1) vector field acquires masses as long as 
it is the centre-of-mass part of the $N$ fractional D3-branes 
of the same type.
Thus, another D-brane has to be introduced, and the U(1) 
will be identified with a linear combination\footnote{The mode 
identified with the inflaton is not necessarily 
the same as the ${\cal N}$ = 2 SUSY partner of the massless U(1) 
vector field.} of the new U(1) and the centre-of-mass U(1) 
\cite{WY04}. 
The D7-branes for SU($M=5$)$_{\rm GUT}$ are wrapped around a 
holomorphic 4-cycle, and chiral multiplets 
in the SU(5)$_{\rm GUT}$-{\bf adj.} representation are usually 
absent in the massless spectrum \cite{IWY,WY01,WY04}. 
Reference \cite{WY04} also discusses how to obtain Higgs particles, 
quarks and leptons from D-branes in curved Calabi--Yau 3-folds.

Note also that 
a model with fractional D3-branes and twisted NS--NS VEV was also 
proposed in \cite{HHK} and one with ordinary D3-branes and 
B-field background in \cite{DHHK}, both 
in the context of hybrid inflation.
The relation between the inflation sector and the sector 
of quarks and leptons in realistic compactification was clearly 
described in \cite{KTW}.

The D-brane construction of the model gives us re-interpretation 
of the results obtained in field theories\footnote{D-brane 
realization is not just a re-interpretation in the context of 
unified theories. It allows us to understand a couple of features 
of the model in a ``unified'' way. 
For more details, see \cite{IWY} or review sections in \cite{WY04}.}. 
With the interpretation in terms of string theory, 
one can now proceed to address the $\eta$-problem, which could 
not be solved within the framework of 
D = 4 ${\cal N}$ = 1 supergravity. 
The assumption 1 is almost trivial, once we have a D-brane model. 
The Planck scale goes to infinity as the volume of a 
Calabi--Yau 3-fold becomes large, 
while the string scale remains finite.
The assumption 2 is also justified, since all the modes 
except D3--D3 open strings are frozen in the large-volume limit.
The inflation sector is now the D3--D7 system on ALE $\times${\bf C} 
in this limit, and hence, 
it has a D = 4 rigid ${\cal N}$ = 2 SUSY (assumption 3).

The rigid ${\cal N}$ = 2 SUSY helps in keeping the potential flat.
The local geometry around the D3--D7 system has to be 
ALE $\times$ {\bf C} so that the ${\cal N}$ = 2 SUSY is preserved. 
The translational invariance in the ${\bf C}$ direction guarantees 
the flatness of the potential at the disc level. 
The effects of stringy-states exchange are at the cylinder level, 
and hence are suppressed by the string coupling constant.  
Moreover, they are well-organized so as not to give rise 
to exponentially growing potential of the inflaton \cite{KTW}.
Since it is mainly from heavy-state exchange amplitudes 
that higher-dimensional operators are expected to be induced,  
the most dangerous effects turn out to be harmless.

Although the above argument clearly shows the benefits 
of the ${\cal N}$ = 2 SUSY and string theory, 
yet it is not enough to justify the assumptions 4 and 5, 
or to guarantee that the potential is sufficiently flat.
When the volume becomes finite, the closed-string modes become 
dynamical, and the inflation sector couples to various 
closed-string modes, not only to the D = 4 gravity mode.
The inflaton potential is generated when closed string moduli 
are integrated out \cite{toward} or when one takes account of 
the effects of twisted Ramond--Ramond field exchange \cite{KTW}, 
and is not sufficiently flat, in general. 
Despite the generic difficulty, it is now clear that we have 
a framework in which gravitational and/or stringy corrections 
are under control and that one is able to see 
whether the potential is sufficiently flat or not. 
Various efforts have been made \cite{HKP,KTW,tye} in this direction, 
and have to be done further.

Let us now turn our attention to the possibility of 
identifying the GUT-breaking model with the model of hybrid 
inflation.
The $\eta$-problem discussed so far is generic to all the cases 
when the vacuum energy is not dominated by D-term. 
Specific to the possibility is that the U(1) gauge-coupling constant 
is not small. 
Analysis in \cite{AHKY} shows that $2 g^2/(4\pi) \lsim 10^{-3}$ 
is necessary for the sufficient e-fold number.
This condition is not satisfied by the GUT-breaking model.
The major cause of the stringent constraint  
is that the inflaton potential in the analysis 
begins to grow exponentially as the field value becomes 
of the order of $M_{\rm pl}$; the inflation should be realized by 
field value sufficiently smaller than $M_{\rm pl}$. 
However, the potential does not grow exponentially 
in the D-brane model, even if the vacuum energy is dominated 
by F-term; this is because of a particular form of the 
K\"{a}hler potential and interaction \cite{Ferrara,KTW}.
Therefore, the upper bound is not applied to the D-brane model, 
and hence, the GUT-breaking model is still eligible to 
be the model of hybrid inflation.

The large coupling constant implies large field value 
at the horizon exit. 
Naive analysis\footnote{There should be no distinction between 
D-term and F-term in the local D-brane model. Thus, the D = 4 
effective action obtained by a simple Kaluza--Klein reduction 
should not have the difference. This is the reason why we consider  
the potential (\ref{eq:Dterm-pot}) to be a (very!) 
crude approximation.} 
using the potential (\ref{eq:Dterm-pot}) shows that the large 
coupling leads to large tensor/scalar ratio 
$r \sim 10^{-2}$ ($\epsilon \sim 10^{-3}$),  
and hence, there is a hope of detecting the tensor mode 
in the CMB polarization.
However, using the potential (\ref{eq:Dterm-pot}) 
is too naive in deriving predictions, since the effects of 
closed-string exchange come in the same order as the 
1-loop correction (as explicitly mentioned in \cite{KTW}).

Finally, there is a remark on the difficulty in the reheating process.
If the inflation sector has an exact rigid ${\cal N}$ = 2 SUSY, 
then the spectrum is completely degenerate, and it is difficult for 
all the particles to decay into lighter particles \cite{WY-hybrid}.
However, the degeneracy is lifted in the extended model based on 
SU($M$)$\times$U($N$) gauge group 
(e.g., \cite{Fujii-Watari,Ibe-Watari}); gauging SU($M$) 
breaks the ${\cal N}$ = 2 SUSY. 
The mass eigenstates of massive SU($N$)$\times$U(1) vector multiplets 
of D = 4 ${\cal N}$ = 1 SUSY are mixtures of those in U($N$) 
and corresponding parts in SU($M$), and this mixing also enables 
the particles to decay \cite{WY-hybrid}. 
Although the reheating process can be understood purely in terms 
of field theories, as above, yet it is remarkable that 
the gauging SU($M$) is an immediate consequence in the D-brane model, 
rather than an additional assumption.

If the hybrid inflation is realized by the GUT-breaking model 
\cite{Izawa-Yanagida}, then one can see what happens in the reheating 
process. The spectrum analysed in \cite{Fujii-Watari,Ibe-Watari} 
tells us that the particles in the inflation sector can decay to 
lighter particles through renormalizable operators, and hence 
the reheating temperature is quite high. 
This allows the thermal leptogenesis \cite{Fukugita-Yanagida} 
to take place, which will be followed 
by late-time entropy production \cite{Fujii-Yanagida,FIY} 
so that the gravitino problem is avoided.

{\bf Note} References \cite{sim,sim2} appeared on the web just before 
we complete this article. They have a little overlap 
with this article.

\section*{Acknowledgements} 

T.W. thanks E.~Gimon, F.~Koyama, and Y.~Tachikawa for discussion.
This work is partially supported by 
the Miller Institute for the Basic Research of Science, 
the Director, Office of Science, Office of High Energy and 
Nuclear Physics, of the U.S. Department of Energy 
under Contract DE-AC03-76SF00098 (T.W.), 
and Grant-in-Aid Scientific Research (s) 14102004 (T.Y.).

\end{document}